\documentclass[a4paper]{article}
\usepackage[dvips]{graphicx}
\usepackage{mathptmx}

\def\abstract{\hfil{\hspace{-13pt}\large{\bf{Abstract}\vspace{-10pt}\\

}}}
\def\bb#1{{\bf{#1}}}
\def\jsum#1#2{\displaystyle\sum_{#1}^{#2}}
\def\jint#1#2{\displaystyle\int_{#1}^{#2}}

\def\jfrac#1#2{\displaystyle\frac{\,#1\,^{\mathstrut}}{\,#2\,_{\mathstrut}}}
\def\jlim#1{\displaystyle \lim_{#1}}
\def\del{\partial}

\def\ee{{\mathrm{e}}}
\newenvironment{e}{\noindent \begin{equation}}{\end{equation}\hspace{-4pt}}
\usepackage{amsmath,amssymb,amsthm}

\newtheorem{thm}{Theorem}

\title{\textbf{A theorem concerning twisted and untwisted partition functions in $U(N)$ and $SU(N)$ lattice gauge theories
\vspace{40pt}}}
\author{Takuya Kanazawa\thanks{Email: tkanazawa@nt.phys.s.u-tokyo.ac.jp}\\
\textit{Department of Physics, University of Tokyo, Tokyo 113-0033, Japan}}
\begin{document}
\date{January 9, 2009}
\maketitle
\begin{abstract}
In order to get a clue to understanding the volume-dependence of vortex free energy (which is defined as the ratio of the 
twisted against the untwisted partition function), we investigate the relation 
between vortex free energies defined on lattices of different sizes. An equality is derived through a simple calculation 
which equates a general linear combination of vortex free energies defined on a lattice to that on a smaller lattice. 
The couplings in the denominator and in the numerator however shows a discrepancy, and we argue that it vanishes in the 
thermodynamic limit. Comparison between our result and the work of Tomboulis is also presented. In the appendix we 
carefully examine the proof of quark confinement by Tomboulis and summarize its loopholes.
\vspace{5pt}
\ \ \ key words: lattice gauge theory, vortex free energy, Migdal-Kadanoff transformation, quark confinement
\end{abstract}
\section{Introduction}
Quark confinement, or (more generally) color confinement is 
one of the most long-standing problems in theoretical physics \cite{AG}. 
So far many proposals have been made concerning the nonperturbative dynamics of QCD which yield confinement, including the 
dual superconductivity scenario \cite{Ripka} and center vortex scenario \cite{vortex} (see \cite{Greensite} for a review), but a 
truly satisfying picture seems to be still missing and precise relationship between different scenarios is elusive. 
In the lattice gauge theory it is formulated as the area law of the Wilson loop in the absence of dynamical fermions, 
which indicates a linear static potential $V(r)\propto r$ between infinitely heavy quark and anti-quark. 
So far the area law has been rigorously proved (in the physicists' sense) 
for quite restricted models \cite{confinement,Yoneya,Mack-Petkova} although it has been numerically checked by 
Monte Carlo simulations for years \cite{Bali}.

The concept of center vortices in non-abelian gauge theories was introduced long time ago 
\cite{vortex,Yoneya,Mack-Petkova,Cornwall}. 
This picture successfully explains many aspects of infra-red properties of Yang-Mills theory at the qualitative 
level, and also at the quantitative level it is reported that 
Monte Carlo simulations show that the value of the string tension can be mostly recovered by the effective 
vortex degrees of freedom (`P-vortex') which one extracts via a procedure called `center projection'
\cite{center_projection}. It is also reported that quenching P-vortices leads to the disappearance of 
area law and the restoration of chiral symmetry {\it{at the same time}} \cite{FD}, which suggests that 
the vortices represent infra-red properties of the theory in a comprehensive manner. 

In addition, the picture that the percolation of center vortices leads to the area law has a firm ground 
based on the Tomboulis-Yaffe inequality \cite{TY}:
\begin{e}
\langle W(C)\rangle\leq 2\Big\{\jfrac{1}{2}(1-\ee^{-F_v})\Big\}^{A_C/L_\mu L_\nu}
\end{e}
for $SU(2)$, which can be proved rigorously on the lattice. Here 
$A_C$ denotes the area enclosed by a rectangle $C$ which lies in a $[\mu,\nu]$-plane, $W(C)$ is the 
Wilson loop associated to $C$ in the fundamental representation and $L_\mu$ is the length of the 
lattice in the $\mu$-th direction. According to this inequality the area law of the l.h.s. follows if the 
vortex free energy $F_v$ vanishes in the thermodynamic limit in such a way that $F_v\approx \ee^{-\rho L_\mu L_\nu}$. 
The quantity $\rho$ gives a lower bound of the string tension, and is called the 't Hooft string tension. This behavior 
of $F_v$ is verified within the strong coupling cluster expansion \cite{Munster} and is also supported by 
Monte Carlo simulations \cite{TK_MC}. So it is worthwhile to study the volume dependence of $F_v$ particularly 
at intermediate and weak couplings. 

In this work we attempt to make a connection between vortex free energies defined on lattices of {\it{different sizes}}.
The set up of this note is as follows. 
In \ref{**} we establish an equality which relates the ratio of the ordinary and the 
twisted partition function on a lattice to that on a smaller lattice, without any restriction on the coupling strength. 
The cost is that there is a 
slight discrepancy between the couplings in the denominator and in the numerator, but it can be shown to tend to zero 
in the thermodynamic limit. In \ref{***} we examine the relation between our result and the work of Tomboulis 
\cite{Tomboulis}.\footnote{Some aspects of Tomboulis' paper which are not touched upon in this note are examined in ref.\cite{Ito-Seiler}.}
Final section is devoted to summary and concluding remarks. 
In the appendix we carefully examine the proof of quark confinement in ref.\cite{Tomboulis}, listing its loopholes.

In view of the serious scarcity of {\it{rigorous}} results in this area of research, our analytical work which involves 
{\it{no}} approximation seems to be of basic importance, and we 
hope that this result will serve as a building block of the proof of quark confinement in the future.

\section{The main result}\label{**}
Let us begin by describing the basic set-up of lattice gauge theory. 
Let $\Lambda$ a $d$-dimensional hypercubic lattice 
of length $L_\mu\,(\mu=1,\dots,d)$ in each direction with the periodic boundary condition imposed. Let 
$\Lambda^{(n)}$ a $d$-dimensional hypercubic lattice of length $L_\mu/b^n\,(\mu=1\dots,d)$ in each direction with a 
 parameter $b\in\mathbf{N}$. The number of plaquettes in $\Lambda$ is denoted by $|\Lambda|$. 
The partition function on the lattice $\Lambda$ is defined as
\begin{e}
Z_\Lambda(\{c_r\})\equiv\jint{}{}\prod_{b\in\Lambda}^{}dU_b\prod_{p\subset\Lambda}^{}f_p(U_p)
\equiv\jint{}{}\prod_{b\in\Lambda}^{}dU_b\prod_{p\subset\Lambda}^{}
\Big[1+\jsum{r\not =\mathbf{1}}{}d_rc_r(\beta)\chi_r(U_p)\Big],
\end{e}
where $dU$ is the normalized Haar measure of gauge group $G$ and $U_p$ is a plaquette variable; $U_p\equiv 
U_{x,\mu}U_{x+\mu,\nu}U_{x+\nu,\mu}^\dagger U^\dagger_{x,\nu}$. In what follows we only consider $G=U(N)$ and $SU(N)$. 
The subscript $r$ labels irreducible 
representations of $G$ and $d_r$ is the dimension, $\chi_r$ is the character of the $r$-th representation 
($\mathbf{1}$ is the trivial representation). 
The coefficients $\{c_r(\beta)\}$ can be determined through the character expansion of (for instance) the Wilson action:
\begin{e}
\ee^{\beta{\mathrm{Re\,Tr\,}}U}\Big/\jint{}{}dU'\,\ee^{\beta{\mathrm{Re\,Tr\,}}U'}
\equiv f_p(U)=1+\jsum{r\not=\mathbf{1}}{}d_rc_r(\beta)\chi_r(U).
\end{e}
It can be checked that $c_r\geq 0$ holds for every $r$, which guarantees the reflection positivity of the measure 
and unitarity of the corresponding quantum-mechanical system.

Multiplying a plaquette variable $U_p$ by a nontrivial element of the center of the gauge group is called a 
{\it{twist}} which, in physical terms, generates a magnetic flux piercing the plaquette. 
Let $C(G)$ denote the center of $G$; $C(SU(N))=Z_N$ and $C(U(N))=U(1)$. The twisted partition function reads
\begin{align}
Z^g_\Lambda(\{c_r\})
&\equiv\jint{}{}\prod_{b\in\Lambda}^{}dU_b
\prod_{p\subset\mathcal{V}}^{}f_p(gU_p)
\prod_{p'\subset\Lambda\setminus\mathcal{V}}^{}f_p(U_{p'}),\hspace{20pt}g\in C(G).
\end{align}
Here $\mathcal{V}$ is a set of stacked plaquettes which winds around $(d-2)$ of the $d$ 
periodic directions of $\Lambda$ forming a $(d-2)$-dimensional torus 
on the dual lattice. (Using reflection positivity one can 
show $Z^{g}_\Lambda\leq Z_\Lambda$ \cite{Kanazawa} and the vortex free energy $F_v^g$ associated to the twist by $g$ 
is defined by $\ee^{-F_v^g}=Z^g_\Lambda/Z_\Lambda$.) Our main result is as follows:
\begin{thm}\label{teiri_}
Let $H$ an arbitrary discrete subgroup of U(1) for $G=U(N)$ and $H=Z_N$ for $G=SU(N)$. 
Fix a set of positive coefficients $\{c_r'\}$ and $n\in\mathbf{N}$, and choose a constant $A^g>0$ 
for each $g\in H$ arbitrarily. Then there exists $\lambda(\{c_r'\})>0$ 
such that for any $0<\alpha<\lambda(\{c_r'\})$ and for 
sufficiently large $|\Lambda|$ there exists $\overline{\alpha}_\Lambda
\equiv\overline{\alpha}_\Lambda(\alpha,\{c_r\},\{c'_r\})$ 
such that 
$\left|\alpha-\overline{\alpha}_\Lambda\right|\leq O\Big(\jfrac{1}{|\Lambda|}\Big)$ and
\begin{e}
\jfrac{\mathcal{Z}_{\Lambda}(\{c_r\})}{Z_{\Lambda}(\{c_r\})}
=\jfrac{\mathcal Z_{\Lambda^{(n)}}(\{\alpha c'_r\})}
{Z_{\Lambda^{(n)}}(\{\overline{\alpha}_\Lambda c'_r\})}     \label{teiri}
\end{e}
with
\begin{e}
\mathcal Z_\Lambda(\{c_r\})\equiv \jsum{g\in H}{}A^gZ^g_\Lambda(\{c_r\}).
\end{e}
\end{thm}
\begin{proof}
Let us define functions $C_\Lambda (\alpha,\{c_r\},\{c'_r\}),\,\mathcal C_\Lambda(\alpha,\{c_r\},\{c'_r\})$ by
\begin{align}
Z_\Lambda(\{c_r\})&=\ee^{C_\Lambda (\alpha,\{c_r\},\{c'_r\})|\Lambda|}Z_{\Lambda^{(n)}}(\{\alpha c'_r\}),\label{47}
\\
\mathcal Z_\Lambda(\{c_r\})&=\ee^{\mathcal C_\Lambda(\alpha,\{c_r\},\{c'_r\})|\Lambda|}\mathcal Z_{\Lambda^{(n)}}(\{\alpha c'_r\}).\label{48}
\end{align}
Taking the ratio of (\ref{47}) and (\ref{48}) yields
\begin{e}
\jfrac{\mathcal Z_\Lambda(\{c_r\})}{Z_\Lambda(\{c_r\})}=\ee^{\{\mathcal C_\Lambda(\alpha,\{c_r\},\{c'_r\})-C_\Lambda (\alpha,\{c_r\},\{c'_r\})\}|\Lambda|}
\jfrac{\mathcal Z_{\Lambda^{(n)}}(\{\alpha c'_r\})}{Z_{\Lambda^{(n)}}(\{\alpha c'_r\})}.\label{773}
\end{e}
From (\ref{773}) and\footnote{$e$ denotes the unit element of $H$.} 
$A^e<\jfrac{\mathcal Z_\Lambda(\{c_r\})}{Z_\Lambda(\{c_r\})},\,\jfrac{\mathcal Z_{\Lambda^{(n)}}(\{\alpha c'_r\})}
{Z_{\Lambda^{(n)}}(\{\alpha c'_r\})}<\jsum{g\in H}{}A^g$ we have
\begin{e}
\left|\mathcal C_\Lambda(\alpha,\{c_r\},\{c'_r\})-C_\Lambda (\alpha,\{c_r\},\{c'_r\})\right|\leq O\Big(\jfrac{1}{|\Lambda|}\Big).
\label{y}
\end{e}
Now let us rewrite (\ref{47}) into more useful form:
\begin{e}
\jfrac{1}{|\Lambda|}\log Z_\Lambda(\{c_r\})
=C_\Lambda (\alpha,\{c_r\},\{c'_r\})+\jfrac{|\Lambda^{(n)}|}{|\Lambda|}\Bigg[
\jfrac{1}{|\Lambda^{(n)}|}\log Z_{\Lambda^{(n)}}(\{\alpha c'_r\})\Bigg].    \label{strong-coupling}
\end{e}
The l.h.s. has a well-defined thermodynamic limit which we denote by $z(\{c_r\})$. 
It is well known \cite{Munster,Kotecky-Preiss} that strong coupling cluster expansion has a non-vanishing radius of 
convergence which is independent of the total volume, so there exists 
$\lambda(\{c_r'\})$ such that $[\dots]$ in the r.h.s. of (\ref{strong-coupling}) can be well approximated by the lowest 
order strong coupling cluster expansion for $0<\alpha<\lambda(\{c_r'\})$, giving
\begin{e}
z(\{c_r\})=C_\infty(\alpha,\{c_r\},\{c'_r\})+\jfrac{a_d}{b^{nd}}\jsum{r\not=\mathbf{1}}{}d_r^2\big(\alpha c_r'\big)^6
+O\big(\big(\alpha c_r'\big)^{10}\big),       \label{--//--}
\end{e}
where $a_3\equiv \jfrac{1}{3},\,a_4\equiv \jfrac{2}{3}$ and we denominated 
$C_\infty(\alpha,\{c_r\},\{c'_r\})\equiv\jlim{|\Lambda|\to\infty}C_\Lambda(\alpha,\{c_r\},\{c'_r\})$. Thus
\begin{e}\hspace{20pt}
\jfrac{dC_\infty(\alpha,\{c_r\},\{c'_r\})}{d\alpha}\simeq -C_0\alpha^5\hspace{30pt}{\rm{for}}\ \ \ 0<\alpha<\lambda(\{c_r'\}),
\end{e}
where the factor $C_0$ is {\it{independent of the volume}}. Therefore $\Big|\jfrac{dC_\Lambda}{d\alpha}\Big|$ has a 
non-vanishing lower bound for arbitrarily large $|\Lambda|$. 
From this fact and (\ref{y}) follows that if 
$|\Lambda|$ is sufficiently large there exists $\overline \alpha_\Lambda(\alpha,\{c_r\},\{c'_r\})$ such that 
$\left|\alpha-\overline{\alpha}_\Lambda\right|\leq O\Big(\jfrac{1}{|\Lambda|}\Big)$ and 
$\mathcal C_\Lambda(\alpha,\{c_r\},\{c'_r\})=C_\Lambda(\overline{\alpha}_\Lambda,\{c_r\},\{c'_r\})$. Hence (\ref{teiri}) is proved.
\end{proof}
One can relax the condition $\{A^g>0\,|\,g\in H\}$ within a scope which does not affect 
the existence of a strictly positive lower bound for 
$\jfrac{\mathcal Z_\Lambda(\{c_r\})}{Z_\Lambda(\{c_r\})}$.

Note that $\alpha$ and $\{c_r'\}$ have no dependence on $\{c_r\}$, for they are parameters we introduced by hand. 
As the free energy density $z(\{c_r\})$ in the l.h.s. of (\ref{--//--}) is a quantity which reflects the phase structure 
of the model, it is not necessarily analytic in $\{c_r\}$ and in $\beta$. In the r.h.s. of (\ref{--//--}), 
such a nonanalyticity in $\{c_r\}$ resides in $C_\infty(\alpha,\{c_r\},\{c'_r\})$ simply because 
the rest of the r.h.s. of (\ref{--//--}) is independent of $\{c_r\}$. Note also that 
$C_\infty(\alpha,\{c_r\},\{c'_r\})$ is Taylor-expandable and analytic in $\alpha$; the analyticity \underline{in $\alpha$} 
and a possible nonanalyticity \underline{in $\{c_r\}$} should be strictly distinguished.
\footnote{The author is grateful to K. R. Ito for valuable correspondence on this point.}

Let us look at (\ref{teiri}) in more detail. Since $\overline{\alpha}_\Lambda$ converges to $\alpha$ in the 
thermodynamic limit, one may be tempted to argue that $\jlim{|\Lambda|\to\infty}
\jfrac{\mathcal{Z}_{\Lambda}(\{c_r\})}{Z_{\Lambda}(\{c_r\})}$ is independent of $\{c_r\}$ because
\begin{e}
\jlim{|\Lambda|\to\infty}
\jfrac{\mathcal{Z}_{\Lambda}(\{c_r\})}{Z_{\Lambda}(\{c_r\})}
=\jlim{|\Lambda|\to\infty}\jfrac{\mathcal Z_{\Lambda^{(n)}}(\{\alpha c'_r\})}
{Z_{\Lambda^{(n)}}(\{\overline{\alpha}_\Lambda c'_r\})}
=\jlim{|\Lambda|\to\infty}\jfrac{\mathcal Z_{\Lambda^{(n)}}(\{\alpha c'_r\})}
{Z_{\Lambda^{(n)}}(\{\alpha c'_r\})}.
\end{e}
It would be valuable to clarify why this claim is incorrect. Let us rewrite (\ref{teiri}) as
\begin{e}
\jfrac{\mathcal{Z}_{\Lambda}(\{c_r\})}{Z_{\Lambda}(\{c_r\})}
=\jfrac{\mathcal Z_{\Lambda^{(n)}}(\{\alpha c'_r\})}
{Z_{\Lambda^{(n)}}(\{\alpha c'_r\})}
\jfrac{Z_{\Lambda^{(n)}}(\{\alpha c'_r\})}
{Z_{\Lambda^{(n)}}(\{\overline{\alpha}_\Lambda c'_r\})}.     \label{neko}
\end{e}
$\left|\alpha-\overline{\alpha}_\Lambda\right|\leq O\Big(\jfrac{1}{|\Lambda|}\Big)$ 
allows us to write $\overline{\alpha}_\Lambda=\alpha+\jfrac{\gamma_{\Lambda}(\alpha,\{c_r\},\{c'_r\})}{|\Lambda^{(n)}|}$ 
with $|\gamma_\Lambda|\leq O(1)$. For $0<\alpha,\,\overline{\alpha}_\Lambda<\lambda(\{c'_r\})$ we may apply 
the convergent cluster expansion to $\log Z_{\Lambda^{(n)}}(\{\alpha c'_r\})$ and 
$\log Z_{\Lambda^{(n)}}(\{\overline{\alpha}_\Lambda c'_r\})$ to obtain
\begin{align}
\jfrac{Z_{\Lambda^{(n)}}(\{\alpha c'_r\})}
{Z_{\Lambda^{(n)}}(\{\overline{\alpha}_\Lambda c'_r\})}
&\equiv\exp\left[|\Lambda^{(n)}|\left(\jfrac{1}{|\Lambda^{(n)}|}\log Z_{\Lambda^{(n)}}(\{\alpha c'_r\})
-\jfrac{1}{|\Lambda^{(n)}|}\log Z_{\Lambda^{(n)}}(\{\overline{\alpha}_\Lambda c'_r\})\right)\right]
\\
&=\exp\left[|\Lambda^{(n)}|\jsum{r,k}{}s_{r,k}\left\{(\alpha c'_r)^k
-(\overline{\alpha}_\Lambda c'_r)^k\right\}\right]
\\
&=\exp\left[-\gamma_\Lambda\jsum{r,k}{}s_{r,k}(c'_r)^kk\alpha^{k-1}+
O\Big(\jfrac{1}{|\Lambda^{(n)}|}\Big)\right]
\\
&\hspace{-5pt}\longrightarrow\ \exp\left[-\gamma_\infty\jsum{r,k}{}s_{r,k}(c'_r)^kk\alpha^{k-1}\right]
\hspace{40pt}\textrm{as}\ |\Lambda|\to\infty.              \label{yumee}
\end{align}
$\{s_{r,k}\}$ are coefficients of clusters. (\ref{yumee}) clearly shows how the discrepancy between $\alpha$ and 
$\overline{\alpha}_\Lambda$ persists in the thermodynamic limit and we now see why the previous claim is incorrect. 
Note, furthermore, that the function $\gamma_\infty(\alpha,\{c_r\},\{c'_r\})
\equiv\jlim{|\Lambda|\to\infty}\gamma_\Lambda(\alpha,\{c_r\},\{c'_r\})$ is not necessarily analytic in $\{c_r\}$; 
information in the l.h.s. of (\ref{neko}) about the phase structure of the model is now packaged within a single 
unknown function $\gamma_\infty$.

We would like to comment on a possible path from theorem \ref{teiri_} to a proof of quark confinement. Suppose $G=SU(N)$.
\begin{thm}
If $N$-ality of the $r$-th representation is non-zero, we have
\begin{e}
|\langle W_r(C)\rangle|\leq 2\Big\{1-\jfrac{1}{N}\jsum{g\in Z_N}{}
\jfrac{Z_\Lambda^g(\{c_r\})}{Z_\Lambda(\{c_r\})}\Big\}^{A_C/L_\mu L_\nu}          \label{ineq}
\end{e}
for the normalized Wilson loop $W_r(C)\equiv \jfrac{1}{d_r}\chi_r\big(\mathcal{P}\prod_{b\in C}U_b\big)$, where $\mathcal P$ implies 
path-ordering and $\langle\dots\rangle$ in the l.h.s. is the expectation value w.r.t. $Z_\Lambda(\{c_r\})$.
\end{thm}
Detailed proof of (\ref{ineq}) is given in ref.\cite{Kanazawa} and we skip it here. 
Let $A^g=\jfrac{1}{N}$ for $^\forall g\in Z_N$ and 
\textit{assume that}\\
\textit{$C_\Lambda(\alpha_0,\{c_r\},\{c'_r\})=\mathcal C_\Lambda(\alpha_0,\{c_r\},\{c'_r\})$ hold for some $\alpha_0\in[0,\lambda(\{c'_r\})]$}. 
Substituting $\alpha =\alpha_0$ into (\ref{773}) gives
\begin{e}
\jfrac{\mathcal Z_\Lambda(\{c_r\})}{Z_\Lambda(\{c_r\})}\Big(\equiv \jfrac{1}{N}\jsum{g\in Z_N}{}
\jfrac{Z_\Lambda^g(\{c_r\})}{Z_\Lambda(\{c_r\})}\Big)
=\jfrac{\mathcal Z_{\Lambda^{(n)}}(\{\alpha_0c'_r\})}{Z_{\Lambda^{(n)}}(\{\alpha_0c'_r\})}.
\end{e}
From the definition of $\lambda(\{c'_r\})$, the r.h.s. can be estimated by 
the convergent cluster expansion, giving
\begin{e}
\jfrac{\mathcal Z_{\Lambda^{(n)}}(\{\alpha_0c'_r\})}{Z_{\Lambda^{(n)}}(\{\alpha_0c'_r\})}\approx 
1-O(\ee^{-\rho L_\mu^{(n)} L_\nu^{(n)}})=1-O(\ee^{-\rho L_\mu L_\nu/b^{2n}})               \label{naisyo}
\end{e}
where $L_\mu^{(n)}\equiv L_\mu/b^n$ is the length of $\Lambda^{(n)}$ in $\mu$-th direction. 
Inserting (\ref{naisyo}) into (\ref{ineq}) we obtain
\begin{e}
|\langle W_r(C)\rangle|\lesssim \ee^{-\rho A_C/b^{2n}},
\end{e}
hence the quark confinement follows. Whether the above strategy (to search for the intersection point of curves 
$\mathcal C_\Lambda(\cdot,\{c_r\},\{c'_r\})$ and $C_\Lambda(\cdot,\{c_r\},\{c'_r\})$) is viable or not remains to be seen.

\section{Comparison with Tomboulis' approach}\label{***}
The purpose of this section is to elucidate the simplification and generalization achieved in the previous section, 
through the comparison with the approach in ref.\cite{Tomboulis}, where only $G=SU(2)$ is treated explicitly. 
The whole argument in ref.\cite{Tomboulis} seems to rest on the Migdal-Kadanoff(MK) renormalization group transformation below:
\begin{e}
c_j(n+1)\equiv \left(\jfrac{\jint{}{}dU\big[f_p(U,n)\big]^{b^{d-2}}\jfrac{1}{d_j}\chi_j(U)}
{\jint{}{}dU\big[f_p(U,n)\big]^{b^{d-2}}}\right)^{b^2r}\in[0,1],\ \ \ \ \ \ \ 
j=\jfrac{1}{2},1,\jfrac{3}{2},\dots             \label{MK}
\end{e}
where $r\in(0,1)$ is a newly introduced parameter and 
\begin{e}
f_p(U,n)\equiv 1+\jsum{j\not=0}{}d_jc_j(n)\chi_j(U),\hspace{10pt}
f_p(U,0)\equiv
\jfrac{\exp{\Big(\jfrac{\beta}{2}{\mathrm{Tr}}\,U\Big)}}
{\jint{}{}dU\,\exp{\Big(\jfrac{\beta}{2}{\mathrm{Tr}}\,U\Big)}},\hspace{10pt}\beta\equiv\jfrac{4}{g^2}.     \label{3++}
\end{e}
When $r=1$, (\ref{MK}) reduces to the original MK transformation \cite{Migdal,Kadanoff}. The reason why $r$ is introduced 
will be briefly explained later. In addition the following quantity is defined:
\begin{e}
F_0(n)\equiv \Big(\jint{}{}dU\,\big[f_p(U,n)\big]^{b^{d-2}}\Big)^{b^2}.
\end{e}
Starting point is the inequality
\begin{e}
Z_{\Lambda}\leq F_0(1)^{|\Lambda^{(1)}|}Z_{\Lambda^{(1)}}(\{c_j(1)\})     \label{111-}
\end{e}
which is proved in appendix A of ref.\cite{Tomboulis}. 
A variable $\alpha\in [0,1]$ and an interpolation function $h(\alpha,t)$ is then introduced, which is supposed to satisfy 
\begin{e}
\jfrac{\del h}{\del \alpha}>0,\ \jfrac{\del h}{\del t}<0,\ h(0, t)=0,\ h(1, t)=1.
\end{e}
The domain of $t\in{\mathbf{R}}$ is arbitrary.

From $1\leq Z_{\Lambda}$ and (\ref{111-}), we see that 
there exists a value $\alpha=\alpha_{\Lambda,h}^{(1)}(t)\in [0,1]$ such that
\begin{e}
Z_{\Lambda}= F_0(1)^{h(\alpha_{\Lambda,h}^{(1)}(t),t)|\Lambda^{(1)}|}
Z_{\Lambda^{(1)}}(\{\alpha_{\Lambda,h}^{(1)}(t)c_j(1)\}).
\label{987}
\end{e}
Similarly it can be shown that there exists a value 
$\alpha=\alpha_{\Lambda,h}^{+(1)}(t)\in [0,1]$ such that
\begin{e}
Z^+_{\Lambda}= F_0(1)^{h(\alpha_{\Lambda,h}^{+(1)}(t),t)|\Lambda^{(1)}|}
Z^+_{\Lambda^{(1)}}(\{\alpha_{\Lambda,h}^{+(1)}(t)c_j(1)\})
\label{988}
\end{e}
where $Z^+\equiv \jfrac{1}{2}(Z+Z^{(-)})$ is introduced for a technical reason
\footnote{The measure of $Z^+$ is reflection positive, which is necessary to derive (\ref{988}). 
The measure of $Z^{(-)}$ is not reflection positive.}
; $Z^{(-)}$ is the twisted partition function for $SU(2)$. 
Note that both (\ref{987}) and (\ref{988}) are independent of $t$. 
Let us take their ratio
\footnote{One can choose different values of 
$t$'s in (\ref{76}) for numerator and denominator, although 
we don't do so here.}:
\begin{align}
\jfrac{Z^+_{\Lambda}}{Z_{\Lambda}}
&=\jfrac{F_0(1)^{h(\alpha_{\Lambda,h}^{+(1)}(t),t)|\Lambda^{(1)}|}
Z^+_{\Lambda^{(1)}}(\{\alpha_{\Lambda,h}^{+(1)}(t)c_j(1)\})}
{F_0(1)^{h(\alpha_{\Lambda,h}^{(1)}(t),t)|\Lambda^{(1)}|}
Z_{\Lambda^{(1)}}(\{\alpha_{\Lambda,h}^{(1)}(t)c_j(1)\})}                               \label{76}
\\
&=\jfrac{F_0(1)^{h(\alpha_{\Lambda,h}^{+(1)}(t),t)|\Lambda^{(1)}|}}
{F_0(1)^{h(\alpha_{\Lambda,h}^{(1)}(t),t)|\Lambda^{(1)}|}}
\jfrac{Z^+_{\Lambda^{(1)}}(\{\alpha_{\Lambda,h}^{+(1)}(t)c_j(1)\})}
{Z^+_{\Lambda^{(1)}}(\{\alpha_{\Lambda,h}^{(1)}(t)c_j(1)\})}
\jfrac{Z^+_{\Lambda^{(1)}}(\{\alpha_{\Lambda,h}^{(1)}(t)c_j(1)\})}
{Z_{\Lambda^{(1)}}(\{\alpha_{\Lambda,h}^{(1)}(t)c_j(1)\})}.
\end{align}
From
\begin{e}
\jfrac{1}{2}<\jfrac{Z^+_{\Lambda}}{Z_{\Lambda}}<1
{\rm{\hspace{20pt}and\hspace{20pt}}}
\jfrac{1}{2}<\jfrac{Z^+_{\Lambda^{(1)}}(\{\alpha_{\Lambda,h}^{(1)}(t)c_j(1)\})}
{Z_{\Lambda^{(1)}}(\{\alpha_{\Lambda,h}^{(1)}(t)c_j(1)\})}<1,
\end{e}
we have
\begin{e}\jfrac{1}{2}<
\jfrac{F_0(1)^{h(\alpha_{\Lambda,h}^{+(1)}(t),t)|\Lambda^{(1)}|}}
{F_0(1)^{h(\alpha_{\Lambda,h}^{(1)}(t),t)|\Lambda^{(1)}|}}
\jfrac{Z^+_{\Lambda^{(1)}}(\{\alpha_{\Lambda,h}^{+(1)}(t)c_j(1)\})}
{Z^+_{\Lambda^{(1)}}(\{\alpha_{\Lambda,h}^{(1)}(t)c_j(1)\})}<2.
\end{e}
This and the fact that 
$Z^+_{\Lambda^{(1)}}(\{\alpha c_j(1)\})$ is a monotonically 
increasing function of $\alpha$ yield
\begin{e}
\left|\alpha_{\Lambda,h}^{(1)}(t)-\alpha_{\Lambda,h}^{+(1)}(t)\right|
\leq O\Big(\jfrac{1}{|\Lambda^{(1)}|}\Big).
\end{e}
This implies that $h(\alpha^{(1)}_{\Lambda,h}(t),t)$ and 
$h(\alpha^{+(1)}_{\Lambda,h}(t),t)$ can be made arbitrarily close 
to each other if one lets $|\Lambda|$ sufficiently large. 
Therefore a slight shift of $t\to t+\delta t$ will enable us to get
\begin{e}
h(\alpha^{(1)}_{\Lambda,h}(t+\delta t),t+\delta t)=h(\alpha^{+(1)}_{\Lambda,h}(t),t),
\end{e}
{\textit{provided that}} $\jfrac{dh(\alpha^{(1)}_{\Lambda,h}(t),t)}{dt}
\not\to 0$ for $|\Lambda|\to\infty$. 
That the newly introduced parameter $r\in(0,1)$ guarantees this can be proved 
through a rather involved calculation; then (\ref{76}) gives
\begin{e}
\jfrac{Z^+_{\Lambda}}{Z_{\Lambda}}
=\jfrac{Z^+_{\Lambda^{(1)}}(\{\alpha_{\Lambda,h}^{+(1)}(t^+)c_j(1)\})}
{Z_{\Lambda^{(1)}}(\{\alpha_{\Lambda,h}^{(1)}(t)c_j(1)\})}.
\end{e}
Repeating above procedure, the following is proved \cite{Tomboulis}:
\begin{thm}\label{teiri*}
For any $n\in{\mathbf{N}}$ and sufficiently large $|\Lambda|$, 
there exist $\alpha_{\Lambda,h}^{(n)}(t_n),\,
\alpha_{\Lambda,h}^{+(n)}(t_n^+)\in[0,1]$ such that\\
$\left|\alpha_{\Lambda,h}^{(n)}(t_n)-\alpha_{\Lambda,h}^{+(n)}(t_n^+)
\right|\leq O\Big(\jfrac{1}{|\Lambda^{(n)}|}\Big)$ 
and
\begin{e}
\jfrac{Z^+_{\Lambda}}{Z_{\Lambda}}
=\jfrac{Z^+_{\Lambda^{(n)}}(\{\alpha_{\Lambda,h}^{+(n)}(t_n^+)c_j(n)\})}
{Z_{\Lambda^{(n)}}(\{\alpha_{\Lambda,h}^{(n)}(t_n)c_j(n)\})}.                  \label{754/}
\end{e}
\end{thm}
It is clear that theorem \ref{teiri*} follows from theorem \ref{teiri_} as a special case 
\Big($H=Z_2,\,c_j'=c_j(n)$ and $A^1=A^{-1}=\jfrac{1}{2}$\Big). 
A difference worth noting is that \textit{the proof of theorem \ref{teiri_} 
necessitates {\it{neither}} the MK transformation (and the related inequality (\ref{111-})) 
{\it{nor}} the special linear combination $Z^+\equiv(Z+Z^{(-)})/2$}. We could entirely 
avoid the complication caused by $r$, which seems to be a significant simplification.

\section{Summary and concluding remarks}
In this note we investigated the lattice gauge theory for general gauge groups with nontrivial center, and 
proved a formula which relates the ratio of twisted and untwisted partition functions to that on the 
smaller lattice. Although the couplings in the numerator and in the denominator cannot be exactly matched, we showed 
the discrepancy to be vanishingly small in the thermodynamic limit. We presented a strategy to prove the 
quark confinement, and also compared our work with Tomboulis' approach in ref.\cite{Tomboulis} clarifying that 
great simplification has occurred in our formulation. 

As has already been clear, our theorem is correct both for $SU(N)$ and for $U(1)$. Whether the theory is 
confining or not, or whether is asymptotically free or not, has nothing to do with the theorem, 
and the same is also true for theorem \ref{teiri*}. Although 
theorem \ref{teiri*} is presented in ref.\cite{Tomboulis} as a cornerstone for the proof of 
quark confinement, it must be confessed that his and our formalism are not quite successful in incorporating the 
dynamics of the theory; entirely new technique might be necessary to prove the quark confinement following the 
strategy described in \ref{**}.

\section*{Acknowledgment}
The author is grateful to Tamiaki Yoneya, Yoshio Kikukawa, Hiroshi Suzuki, Shoichi Sasaki, Shun Uchino, Keiichi R. Ito 
and Terry Tomboulis for useful discussions. 
He also thanks Tetsuo Hatsuda and Tetsuo Matsui who stimulated him to investigate ref.\cite{Tomboulis}. 
This work was supported in part by Global COE Program 
``the Physical Sciences Frontier'', MEXT, Japan.

\appendix
\section{Appendix}
In the following we comment on the validity of the advocated proof of quark confinement in 
four dimensional $SU(2)$ lattice gauge theory \cite{Tomboulis}, to help the reader understand 
the precise relation between our work and ref.\cite{Tomboulis}. The reader who is only interested in a qualitative 
understanding on the viability of the proposed proof may want to go directly to the last paragraph of this appendix, 
in which a less technical, quite intuitive overview of the situation is presented.
\subsection{Correctness}
First of all, the proof is mathematically incomplete at least in \textit{four} aspects. 
\begin{enumerate}
\item 
In theorem \ref{teiri*} of this paper, 
$\alpha_{\Lambda,h}^{+(n)}(t_n^+)$ numerically differs from $\alpha_{\Lambda,h}^{(n)}(t_n)$ in general. However, \textit{if} 
we could find $t^*\in\mathbf{R}$ (at least for $n\gg 1$ for which $c_j(n)\ll 1$; we will discuss this point shortly) 
such that $\alpha_{\Lambda,h}^{+(n)}(t^*)=\alpha_{\Lambda,h}^{(n)}(t^*)$, then 
the usual strong coupling cluster expansion method could be applied to the r.h.s. of (\ref{754/}), thus giving 
a proof of quark confinement. In this way Tomboulis reduced the problem of quark confinement 
to the problem of showing the existence of such $t^*\in\mathbf{R}$. Furthermore he attempted to find such $t^*$ 
by interpolating $\alpha_{\Lambda,h}^{+(n)}(t)-\alpha_{\Lambda,h}^{(n)}(t)=0$ (the equation to be solved) 
and another independent equation whose solution we know does exist, 
by one parameter $0\leq \lambda\leq 1$. The result is a single two-variable equation 
$\Psi(\lambda,t)=0$, where $\Psi(1,t)=0$ is equivalent to $\alpha_{\Lambda,h}^{+(n)}(t)-\alpha_{\Lambda,h}^{(n)}(t)=0$ 
and $\Psi(0,t)=0$ is the other equation that can be solved by some $t_0\in\mathbf{R}$. Assuming the existence of 
$t(\lambda)$ such that $\Psi(\lambda,t(\lambda))=0$ and then differentiating both sides by $\lambda$, we have
\begin{e}
\jfrac{dt}{d\lambda}=-\jfrac{\Psi_{,\lambda}(\lambda,t)}{\Psi_{,t}(\lambda,t)}   \label{yume_}
\end{e}
where the subscripts denote partial derivatives. ($\Psi_{,t}(\lambda,t)\not =0$ is assumed in (\ref{yume_}), but it 
is rigorously proved in ref.\cite{Tomboulis}.) Thus what we should do is to to solve the differential equation 
(\ref{yume_}) with the initial condition $t(0)=t_0$ and find $t(1)$, which is nothing but $t^*$. 
Here comes the crux of the problem: he argues that $\Psi_{,t}(\lambda,t)\not =0$ for $0\leq \lambda\leq 1$ is a 
sufficient condition for the existence of $t(1)$ since one can extend $t(\lambda)$ from $t(0)$ to $t(1)$ by 
iteratively integrationg both sides of (\ref{yume_}). 
But \textit{it is not the case}. Indeed one can find infinitely many counterexamples as follows. 
Take arbitrary functions $f(t),\,g(t),\,t\in\mathbf{R}$ such that 
both are strictly increasing (or both decreasing) and $f(t)=0$ has \textit{no} solution while $g(t)=0$ has a solution. 
Then $\Psi(\lambda,t)=\lambda f(t)+(1-\lambda)g(t)$ satisfies 
$\Psi_{,t}\not=0$ for all $0\leq \lambda\leq 1$ and yet $t(1)$ does not exist.\footnote{This 
flaw was first pointed out by the author and was recapitulated in ref.\cite{Ito-Seiler}.} Thus the task of showing 
the existence of $t(1)(=t^*)$ is highly nontrivial.
\item 
The second point concerns the behavior of $\{c_j(n)\}$ for $n\gg 1$. We have already noted that the definition (\ref{MK}) 
of $\{c_j(n)\}$ is different from the original MK transformation by an additional parameter $0<r<1$. It has been 
rigorously proved \cite{Ito,Muller-Schiemann} for the original MK transformation ($r=1$) 
that the effective coupling ($\beta$) necessarily flows to the strong
coupling limit ($\beta=0$) after sufficiently many iterations,
regardless of the initial coupling and regardless of the gauge group as long as it is compact. Hence it is natural to ask 
whether this property also holds for the modified MK (MMK) transformation ($r<1$) or not.

A simple strong-coupling calculation for $SU(2)$ shows that through one MMK transformation 
the effective coupling changes from $\beta$ to $\beta^{b^2r}$;
for $\jfrac{1}{b^2}<r<1$ we have $\beta>\beta^{b^2r}$ and the flow goes toward the strong coupling limit, while 
for $0<r<\jfrac{1}{b^2}$ we have $\beta<\beta^{b^2r}$ and the flow goes toward the \textit{weak coupling limit}. 
This result is depicted in fig.\ref{0}.\footnote{We obtained an analogous result for $U(1)$.} 
We are therefore left with only three distinct possibilities; see fig.\ref{1},\ \ref{2} and \ref{3}.
\begin{figure}[hbtp]
\begin{minipage}{.45\textwidth}
\begin{center}
\includegraphics[width=4.0cm,clip]{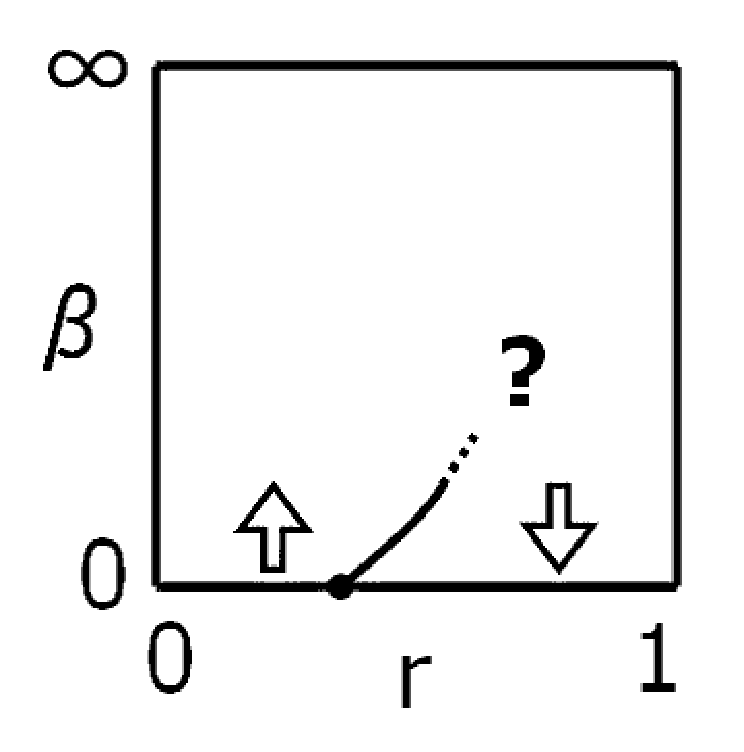}
\end{center}\vspace{-25pt}
\caption{A schematic flow diagram deduced from a strong coupling expansion.}
\label{0}
\end{minipage}
\hfill
\begin{minipage}{.45\textwidth}
\begin{center}
\includegraphics[width=4.0cm,clip]{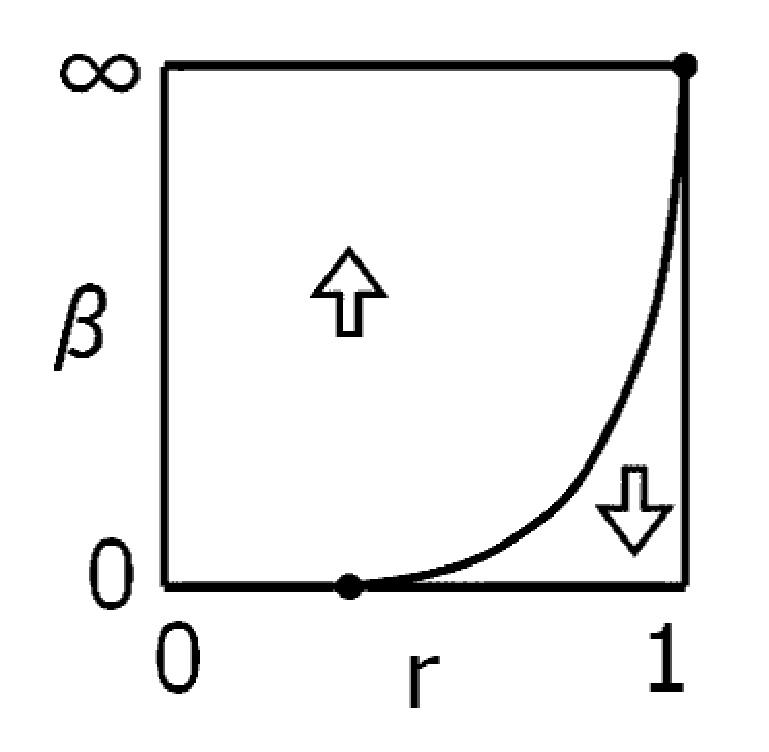}
\end{center}\vspace{-25pt}
\caption{A possible extrapolation. The critical line ends at $(r,\beta)=(1,\infty)$.}
\label{1}
\end{minipage}
\vspace{15pt}
\\
\begin{minipage}{.45\textwidth}
\begin{center}
\includegraphics[width=4.0cm,clip]{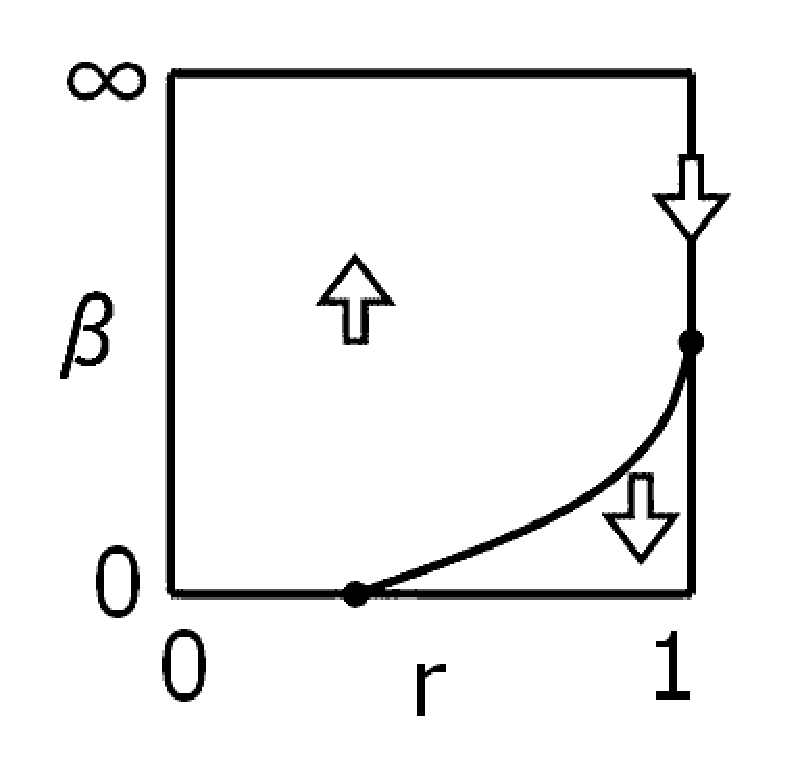}
\end{center}\vspace{-25pt}
\caption{A possible extrapolation. The critical line ends somewhere on the line $r=1$.}
\label{2}
\end{minipage}
\hfill
\begin{minipage}{.45\textwidth}
\begin{center}
\includegraphics[width=4.0cm,clip]{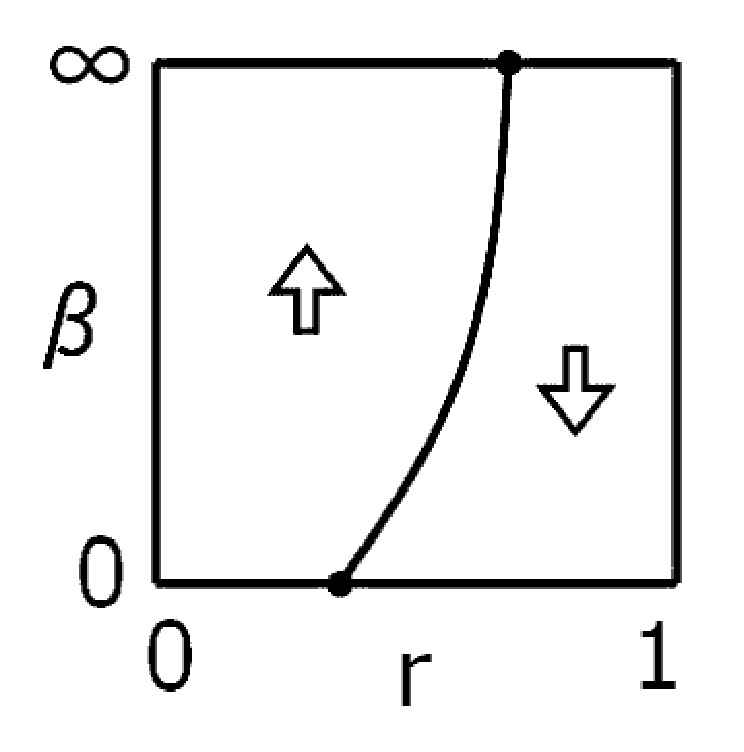}
\end{center}\vspace{-25pt}
\caption{A possible extrapolation. The critical line ends somewhere on the line $\beta=\infty$.}
\label{3}
\end{minipage}
\end{figure}
However, a simple weak-coupling calculation for $SU(2)$ shows that by one MMK transformation the effective coupling changes 
from $\beta$ to $\beta/r\ (>\beta)$.\footnote{Again, we obtained an analogous result for $U(1)$.} So the flow goes 
toward the weak couling limit for sufficiently large $\beta\gg 1$ and for any \textit{fixed} $r<1$, 
which implies that we can exclude the possibility of fig.\ref{3}. At present we do not know 
which of fig.\ref{1} or \ref{2} is the correct one. 

But does such a difference matter? 

Yes, it \textit{strongly} does. For illustration, let us believe fig.\ref{2}. Then for \textit{any} $r<1$ and large 
initial coupling $\beta\gg 1$, the flow would go to the weak coupling limit: $\jlim{n\to\infty}c_j(n)=1$. 
In such a case, the existence of $t(1)=t^*$ no longer guarantees quark confinement, since the strong coupling 
cluster expansion is inapplicable to the r.h.s. of (\ref{754/}). Thus the alleged proof completely fails for $\beta\gg 1$
in the situation of fig.\ref{2}. In contrast, if fig.\ref{1} were the case, then for arbitrary initial $\beta$ 
we can always find an appropriate value of $r<1$ for which the flow still goes to the strong coupling limit, hence 
the proof is valid (except for the step of showing the existence of $t^*$). 

These considerations clearly show the importance of ascertaining the qualitative flow of coupling under 
the MMK transformation, 
but Tomboulis in ref.\cite{Tomboulis} simply asserts that $r<1$ can be chosen very close to 1 so that the flow to the 
strong coupling limit is not destroyed. (This amounts to assuming the situation of fig.\ref{1} implicitly.) 
However he gave neither 
analytical nor numerical evidence of his claim. This constitutes the second incomplete point of the proof.

As an aside we note that it is not mandatory to use a fixed value of $r$ throughout a flow. For example, for a 
fixed initial coupling $\beta$, we are allowed to choose 
different values of $r$ for every step of decimation, so that $\{c_j(1)\}_j$ is calculated from 
$\{c_j(0)\}_j$ via (\ref{MK}) with $r=r_1<1$, $\{c_j(2)\}_j$ is calculated from $\{c_j(1)\}_j$ via 
(\ref{MK}) with $r=r_2<1$, ... and so on. However we can verify it does not make much difference: indeed, 
if fig.\ref{2} is true and $\beta\gg 1$, the flow would inevitably go to the weak coupling limit ($\beta\to\infty$) 
regardless of whatever values we choose for $\{r_i\}_i$. It is still vital even under such a looser condition 
to find out the flow diagram of the MMK transformation.
\item 
The third criticism\footnote{It has already been discussed at length in ref.\cite{Ito-Seiler}.} 
is more indirect than the two given above: since the compact $U(1)$ lattice gauge theory in four dimensions 
has a deconfined phase at weak coupling \cite{Guth-Flohrich-Spencer}, 
the proposed proof must fail for the abelian theory, but \textit{which specific step of the proof fails?} 

Logically speaking, there are only two ways to reply:
\begin{enumerate}
\item ``The difference between abelian and nonabelian theories does not emerge until we complete the 
proof of the existence of $t^*=t(1)$. Such a proof, if any, is expected to become rather involved, 
since the delicate difference between abelian and nonabelian theories on the lattice has to be accounted for.''
\item ``The flow of the effective coupling under the MMK transformation is significantly different for abelian and 
nonabelian theories. Fig.\ref{2} applies to the abelian case, while fig.\ref{1} applies to the nonabelian case, 
which implies that the proof surely fails for the abelian case.\footnote{Later we will see that this is not the whole story.}
''
\end{enumerate}
Much more work will be needed to determine which scenario is the right one.
\item 
The fourth point concerns the necessity of the parameter $r$ itself. Before in this article we briefly explained 
the reason why $r$ was introduced; it was to avoid $\jfrac{dh(\alpha^{(n)}_{\Lambda,h}(t),t)}{dt}
\to 0$ as $|\Lambda|\to\infty$.\footnote{It would be worth noting that 
how close $r$ is to 1 does not matter for this purpose: it is whether $r=1$ or $r<1$ that counts.} 
However it was not shown in ref.\cite{Tomboulis} whether such an undesirable behavior \textit{actually} occurs or not. 
It is easy to see that it will occur if and only if the original MK transformation ($r=1$) becomes exact in the 
thermodynamic limit insofar as the free energy density is concerned. However, considering that 
the MK transformation is an approximate decimation scheme 
which bunches $b$ parallel adjacent plaquettes into one, it seems more plausible that the MK transformation 
can be exact {\it{at zero coupling}} ($\beta=\infty$) \textit{alone}; in such a case, 
any fluctuation disappears and the actions of adjacent plaquettes will have equal expectation values, thus 
making the bunching an exact procedure. 

So the author's guess is that $r$ is {\textit{unnecessary}} to establish theorem \ref{teiri*}. 
If it were the case, we could dispense with any modification of the original MK transformation. It is a sad news 
for the whole proof, however, since the abelian and nonabelian theories qualitatively behave in the same way 
under the original MK transformation (recall that the effective coupling flows to the strong coupling limit 
under the iterated MK transformations regardless of the initial coupling 
\textit{whether the gauge group is abelian or nonabelian}) and 
so the difference between abelian and nonabelian theories has to be explained 
within the proof of the existence of $t^*=t(1)$, implying that much remains to be done to prove the 
quark confinement.

Though our guess is by no means conclusive, this point deserves further study.
\end{enumerate}
These are the main aspects of the alleged proof which are not clear enough to the author. The fact is that, 
the ultimate goal of ref.\cite{Tomboulis} (giving a proof of quark confinement) is not yet achieved, 
and instead, what has already been actually 
proved with a satisfactory rigor is theorem \ref{teiri*}. One purpose of this paper is to point out 
that it can be derived in a far simpler way, bypassing the complicated discussion on the MMK transformation at all.

\subsection{Discussion}
First we state and prove a theorem. Its intriguing implication will be described afterwards in relation to 
our previous analyses.
\begin{thm}\label{theorem_****}
Let the gauge group (denoted by $G$) a compact and connected Lie group. 
With $r\equiv 1-\jfrac{1}{b^2}$, the coefficients 
$\{c_j(n)\}$ defined by (\ref{MK}) converge to the strong-coupling limit \big(i.e. 
$\jlim{n\to\infty}c_j(n)=0$\big) if $b\in{\mathbf{N}}$ is chosen sufficiently large, 
depending on the initial coupling $\{c_j(0)\}$.
\end{thm}
\begin{proof}
Let us define a functional space
\begin{e}
\mathfrak{F}\equiv\{f\in C^2(G)\,\big|\,f>0,\,f({\mathbf{1}})=1\}
\end{e}
(${\mathbf{1}}$ is the unit element of $G$ and $C^2(G)$ is a set of continuously twice-differentiable functions on $G$) 
and a map $T:\mathfrak{F}\to\mathfrak{F}$ by
\begin{e}
Tf(U)\equiv\jfrac{(f^q)^{*s}(U)}{(f^q)^{*s}({\mathbf{1}})},\ \ \ \ \ s,q\in{\mathbf{N}},\ s\geq 2,\ U\in G,\ 
f\in\mathfrak{F}.     \label{---}
\end{e}
Here $(\dots)^{*s}$ denotes an $s$-fold convolution. 
(\ref{---}) is nothing but the original MK transformation ($r=1$); if we put $q=s=b^2$, 
(\ref{---}) reduces to (\ref{MK}) and (\ref{3++}) with $r=1$ except for an irrelevant normalization factor. 
Note that $Tf$ is a class function over $G$ if $f$ is. Let $V_f(U)\equiv -\log f(U)$.

Introducing a specific functional $W[f]$ (whose explicit form is not necessary for our purpose), 
$\rm{M\ddot uller}$ and Schiemann \cite{Muller-Schiemann} proved the following inequalities which 
hold for $^\forall f\in\mathfrak{F}$:
\begin{gather}
(V_f)_{\mathrm{max}}-(V_f)_{\mathrm{min}}\leq\jfrac{1}{2}W[f],  \label{/6}
\\
W[Tf]\leq\jfrac{q}{s}(1-e^{-W[f]})W[f],   \label{/7}
\end{gather}
where $(V_f)_{\mathrm{max}}$ \big($(V_f)_{\mathrm{min}}$\big) is the maximum (minimum) value of $V_f$ over $G$. 
(We would like to emphasize that \textit{the definition of $W[f]$ and the property (\ref{/6}) are entirely 
independent of the map $T$.}) We refrain from reproducing their proof here. Instead we would like to 
note one of its major consequences. 
If $\jfrac{q}{s}\leq 1$ holds, (\ref{/7}) implies that 
$W[T^nf]\leq W[f]$ for $^\forall n\in\mathbf{N}$. From this and (\ref{/6}),\ (\ref{/7}) we obtain
\begin{gather}
\hspace{50pt}  0\leq W[T^nf]\leq\big(1-e^{-W[f]}\big)^nW[f],\ \ \ \ \ \ \ \ {^\forall n\in\mathbf{N}}.
\\
\hspace{-15pt}\therefore \ \jlim{n\to\infty}W[T^nf]=0.     \label{zero}
\end{gather}
(\ref{/6}) and (\ref{zero}) imply that $\jlim{n\to\infty}V_{T^nf}$ is a constant function on $G$ 
and so is $\jlim{n\to\infty}T^nf$. Thus, if $f$ is a class function on $G$, 
every coefficient in the character expansion of $T^nf$ would tend to $0$ as $n\to\infty$ except for a constant term. 
This completes the proof of the convergence of the \textit{original} MK transformation. 

Let us then prove theorem \ref{theorem_****}. 
If we introduce $r\in(0,1)$, $s$ changes from $b^2$ to $b^2r$ while $q=b^2$ remains unchanged. However 
we have to be careful at this point: $b^2r$ is not integral in general while (\ref{/6}) and (\ref{/7}) have been proved 
only for an integral $s$. Let us choose, for example, $r=1-\jfrac{1}{b^2}$ so that $b^2r\in{\mathbf{N}}$. 
Then (\ref{/7}) yields
\begin{e}
W[Tf]\leq\jfrac{1-e^{-W[f]}}{r}W[f].
\end{e}
Let us take $b$ so large that $\jfrac{1-e^{-W[f]}}{r}<1$. Then we have $W[T^nf]\leq W[f]$ 
also in this case. It is straightforward to derive
\begin{gather}
\hspace{55pt}  0\leq W[T^nf]\leq\left(\jfrac{1-e^{-W[f]}}{r}\right)^nW[f],\ \ \ \ \ \ \ \ {^\forall n\in\mathbf{N}}.
\\
\hspace{-15pt}\therefore \ \jlim{n\to\infty}W[T^nf]=0.
\end{gather}
This completes the proof of theorem \ref{theorem_****}.
\end{proof}
The most important feature of theorem \ref{theorem_****} is that it holds both for abelian and for nonabelian 
theories. In the third discussion of the previous section, we pointed out the possibility that 
the flow of the effective coupling under the MMK transformation for abelian theories might be significantly different 
from that of nonabelian theories making consistent the alleged proof of quark confinement for $SU(2)$ and the deconfinement 
phase of the $U(1)$ theory mutually. However, surprisingly, theorem \ref{theorem_****} has made it clear 
that for any initial coupling we can choose $r$ smaller than 1 without messing the renormalization group 
flow toward the strong coupling limit, \textit{both for abelian and for nonabelian theories.} 
The price we have to pay is just to choose an appropriate value of $b\in\mathbf{N}$, which does the proof no harm, 
since none of the steps of the proof explicitly depends on the choice of $b$. Thus, qualitatively speaking, there are 
only two scenarios possible:
\begin{enumerate}
\item ``
We should prove the existence of $t^*=t(1)$ in such a way that does not permit to choose $b\gg 1$ for $\beta\gg 1$.''
\item ``
We should prove the existence of $t^*=t(1)$ in such a way that incorporates the nonperturbative dynamics of the theory 
well enough to reveal the delicate difference between abelian and nonabelian theories on the lattice.''
\end{enumerate}
If the first scenario were to be realized, then the prescription learned from theorem \ref{theorem_****} (to vary $b$ 
depending on the initial coupling) is not implementable and we again face with the need to understand the flow 
diagram of the MMK transformation. But the first scenario looks quite bizarre in the sense that any argument 
that is valid for, say, $b\leq 3$ but becomes completely invalid for $b>4$ seems to be unphysical.

Thus we regard the second scenario as the one to be pursued. Since it essentially says that 
the proposed proof in its present form is unable to tell the abelian from the nonabelian theory at all, 
we are inclined to conclude that 
\textit{the proposed proof has made little or no progress compared with the original MK transformation.} 
\\

We would like to summarize our comments on the present situation. 
The main tool exploited in ref.\cite{Tomboulis} 
is the Migdal-Kadanoff transformation (with a modification). It is an \textit{approximation}; it is not the genuine, 
exact renormalization group transformation of the Yang-Mills theory on the lattice. Therefore if one is to make use of 
it to say something exact about the \textit{real} Yang-Mills theory, it could be done \textit{only after 
exactly understanding the relation between the Migdal-Kadanoff approximation and the Yang-Mills theory.} 
However it does not seem to be done in ref.\cite{Tomboulis}; although a rigorous inequality is proved concerning 
their relation, it is generally impossible to obtain an exact value via an inequality 
(an inequality is not an equality). 
Suppose we want to solve the equation $f(x)=0$ with the knowledge that the solution surely exists in the range $2<x<3$. 
By interpolation it can be rewritten as $x=2c+3(1-c)$ for some $0<c<1$. It is 
in the disguise of an \textit{equality}, 
but we know that no new information is obtained through an interpolation alone. 
Such interpolations are repeatedly used throughout ref.\cite{Tomboulis}, but obviously they give us no new 
information about confinement. Essentially this is the very reason the approach of ref.\cite{Tomboulis} is 
unsuccessful and looks hard to remedy.

\end{document}